\documentclass[11pt]{article} % use larger type; default would be 10pt

\usepackage[colorlinks=true, allcolors=blue]{hyperref}

\usepackage[utf8]{inputenc} % set input encoding (not needed with XeLaTeX)

\usepackage{geometry} % to change the page dimensions
\usepackage{graphicx} % support the \includegraphics command and options

\usepackage[round]{natbib}

\usepackage{booktabs} % for much better looking tables
\usepackage{array} % for better arrays (eg matrices) in maths
\usepackage{paralist} % very flexible & customisable lists (eg. enumerate/itemize, etc.)
\usepackage{verbatim} % adds environment for commenting out blocks of text & for better verbatim
\usepackage{subfig} % make it possible to include more than one captioned figure/table in a single float
\usepackage{amsmath}
\usepackage{amsfonts}
\usepackage{amssymb}
\usepackage{mathrsfs}
\usepackage{bbold}

\usepackage{multirow}

\allowdisplaybreaks

\newcommand{\dd}{\mathrm{d}}
\newcommand{\FIRM}{\mathrm{FIRM}}

\newcommand{\MCB}{\mathrm{MCB}}
\newcommand{\DSC}{\mathrm{DSC}}
\newcommand{\UNC}{\mathrm{UNC}}

\usepackage{fancyhdr} % This should be set AFTER setting up the page geometry
\usepackage{sectsty}
\allsectionsfont{\sffamily\mdseries\upshape} % (See the fntguide.pdf for font help)
\usepackage[nottoc,notlof,notlot]{tocbibind} % Put the bibliography in the ToC
\usepackage[titles,subfigure]{tocloft} % Alter the style of the Table of Contents

\title{A User-Focused Approach to Evaluating Probabilistic and Categorical Forecasts}

\author{Nicholas Loveday, Robert Taggart, Mohammadreza Khanarmuei\\Australian Bureau of Meteorology\\nicholas.loveday@bom.gov.au}

\begin{document}

\maketitle

\begin{abstract}
\noindent A user-focused verification approach for evaluating probability forecasts of binary outcomes (also known as probabilistic classifiers) is demonstrated that is (i) based on proper scoring rules, (ii) focuses on user decision thresholds, and (iii) provides actionable insights. It is argued that when categorical performance diagrams and the critical success index are used to evaluate overall predictive performance, rather than the discrimination ability of probabilistic forecasts, they may produce misleading results. Instead, Murphy diagrams are shown to provide better understanding of overall predictive performance as a function of user probabilistic decision threshold. It is illustrated how to select a proper scoring rule, based on the relative importance of different user decision thresholds, and how this choice impacts scores of overall predictive performance and supporting measures of discrimination and calibration. These approaches and ideas are demonstrated using several probabilistic thunderstorm forecast systems as well as synthetic forecast data. Furthermore, a fair method for comparing the performance of probabilistic and categorical forecasts is illustrated using the FIxed Risk Multicategorical (FIRM) score, which is a proper scoring rule directly connected to values on the Murphy diagram. While the methods are illustrated using thunderstorm forecasts, they are applicable for evaluating probabilistic forecasts for any situation with binary outcomes.
\vspace{10pt}

\noindent\textbf{Keywords:} forecast verification; probabilistic forecasting; categorical forecasts, proper scoring rule, performance diagrams, weather forecasting, thunderstorms.
\end{abstract}

\noindent{\textit{This Work has been submitted to Weather and Forecasting. Copyright in this Work may be transferred without further notice.}}

%%%%%%%%%%%%%%%%%%%
\section{Introduction}
Probabilistic weather forecasts are commonly produced by meteorological agencies and have existed since at least the eighteenth century \citep{murphy1998early}. They are commonly expressed as the probability of binary outcome occurring (e.g., an 80\% chance of rainfall tomorrow), with such forecasts sometimes called ``probabilistic classifiers''. Evaluation of weather forecasts is critical to understand their quality, behavior, accuracy, and value. In the middle of the twentieth century,  the Brier score \citep{brier1950verification} was introduced and has become one of the most commonly used verification metrics to evaluate probabilistic forecasts of binary outcomes. Whilst many of the classical evaluation tools developed since then are known to the meteorological community (e.g., reliability diagrams), recent advances published in the statistics literature are not well-known. A primary aim of this paper is to review these advances and promote their use.

The Australian Bureau of Meteorology (hereafter, ``the Bureau'') recently started routinely producing daily probability of thunderstorm (DailyPoTS) forecasts using a variety of techniques and wanted their quality assessed. This prompted us to review the verification methods used to evaluate probabilistic thunderstorm forecasts in the literature. Over the last two decades there has been a surge of probabilistic and categorical thunderstorm forecast systems produced in research and operational environments \citep{bright2005physically, bright2009short, dance2010thunderstorm, craven2018overview, simon2018probabilistic, rothfusz2018facets, charba2019lamp, brunet2019evaluation, brown2019objective, harrison2022utilizing}. The methods used to evaluate probabilistic thunderstorm forecasts in these systems include the Brier score \citep{brier1950verification} and the corresponding Brier skill score, receiver operating characteristic (ROC) curves, reliability diagrams, the Pierce skill score, relative economic value \citep{richardson2000skill}, and categorical performance diagrams \citep{roebber2009visualizing}. The latter displays probability of detection (POD), success ratio (1 - false alarm ratio) (SR), critical success index (CSI), also known as the threat score, and frequency bias (FB). These methods are also commonly used to verify other forecast parameters such as a ``chance of rainfall'' forecast. While there have been many approaches taken to verify forecasts of binary outcomes in the meteorological literature, we suggest that best practice should meet three key requirements.

\begin{enumerate}
\item Scores used to assess overall predicitive performance should be proper scoring rules within a probabilistic forecast framework \citep{winkler1968good, gneiting2007strictly}.  A scoring rule assigns a number to a forecast--observation pair that is indicative of predictive performance. A scoring rule is proper if, from the forecaster's perspective, their expected score will be optimized by issuing the forecast that corresponds to their true belief. Using a proper scoring rule 
ensures that forecasters or forecast system developers avoid facing the quandary of either issuing an honest forecast or of issuing a different forecast that will optimize the verification score. This quandary is known as the ``forecaster's dilemma" \citep{lerch2017forecaster} and choosing to optimize the score in such circumstances is sometimes referred to as ``hedging''. In the case of issuing probabilistic forecasts for binary outcomes, proper scoring aligns with the notion of consistency outlined in Murphy's famous essay on what constitutes a good forecast \citep{murphy1993good} and expanded on by \citet{gneiting2011making}. Examples of proper scoring rules are the Brier 
score, the logarithmic score \citep{good1952rational}, and the misclassification rate, which is widely used in machine 
learning.
\item User-focused verification should measure performance across important decision thresholds \citep[e.g.,][]{rodwell2020user, foley2020comparison, taggart2022evaluation, laugesen2023flexible}. In the case of probabilistic forecasts of binary outcomes, a decision threshold is the forecast probability value that a user will take action on. E.g., a city will cancel their New Years Eve fireworks display if the forecast chance of a thunderstorm exceeds 60\%. Using such simple decision models, the Brier score weights all user decision thresholds equally \citep[c.f.][]{shuford1966admissible, schervish1989general}, 
while the logarithmic score heavily weights decision thresholds closer to 0 and 100\%. Different proper scores may rank the performance of different forecast systems differently. When comparing forecast systems, it is critical that one understands performance for different decision thresholds and that the chosen proper scoring rule weights each decision threshold appropriately.
\item Methods that provide actionable insights into the behavior of a forecast system, such as measures of conditional calibration
 or sharpness, should be used to complement overall predictive performance scores. Insights on calibration can inform where meteorologists, forecast system developers and users of forecasts can provide bias corrections. Measures of discrimination can highlight which forecast system has the highest potential predictive ability subject to recalibration.
\end{enumerate}

In this paper we demonstrate that using categorical performance diagrams to rank the overall predictive performance of probabilistic forecast systems can lead to misguided inferences, introduce forecaster's dilemmas and hence does not meet requirement 1 listed above for overall predictive performance measures. Instead they may be used to infer discrimination ability and support requirement 3 listed above. Even when categorical performance diagrams are presented alongside reliability diagrams, the ranking of overall predictive performance of forecast systems may still be unclear. On the other hand, miscalibration--discrimination diagrams and Murphy diagrams are visual tools from which overall predictive performance can be inferred \citep{ehm2016quantiles, dimitriadis2023evaluating}. Furthermore, we demonstrate how recent advances in generating reliability and ROC diagrams support requirement 3 listed above. We demonstrate these tools by comparing three forecast systems and products that predict thunderstorm probability over the Australian region. 

In addition to DailyPoTS forecasts, the Bureau also issues categorical thunderstorm outlook products generated by operational meteorologists, where the categories correspond to different likelihoods of thunderstorm occurrence. We demonstrate that the FIxed Risk Multicategorical (FIRM) score \citep{taggart2022scoring} permits a fair comparison between such categorical forecasts with probabilistic forecasts.

While the methods in this paper are illustrated using thunderstorm forecasts, they are applicable for evaluating probabilistic forecasts for any situation with binary outcomes.

\section{Data} \label{data}
\subsection*{Forecast data} 
This study uses forecasts from the 2022-2023 Australian summer (December-February). The Bureau issues gridded (approximately 6km grid spacing) DailyPoTS forecasts twice a day. These gridded forecasts are used to populate the weather forecasts on the Bureau's website and mobile phone app. In the future, these thunderstorm grids may also be made available for downstream users. The forecasts are issued for seven lead days (1-7) and are valid for 24-hour periods that begin at 15 UTC, which is 30 minutes from local midnight Australian Central Standard Time. The forecasts are defined as the likelihood that at least one lightning strike (either cloud-to-ground or cloud-to-cloud) will be observed within a 10-km radius of the center of a grid cell.

\subsection*{AutoFcst}
AutoFcst is an automated forecast system that takes post-processed output from post-processed consensus forecasts as well as a global wave model and produces a coherent set of forecast grids for many parameters \citep{griffiths2022autofcst}. For thunderstorm forecasts, AutoFcst utilizes probabilistic predictions from the Bureau's operational lightning guidance system, Calibrated Thunder (CalTS), which is a modified version of the system described by \citet{bright2005physically}.  CalTS combines ensemble NWP model output with historical lightning observations to produce calibrated forecasts of the probability of one or more lightning strikes within 10 km of a point across Australia and surrounding coastal waters. In this paper we use the 12 UTC base time run of AutoFcst, which makes use of the 12 UTC run of CalTS. A lead day 1 forecast starts 27 hours after the base time and approximately 20 hours after it is available for operational meteorologists to use.

\subsection*{New Calibrated Thunder Guidance}
A hindcast of a research version of CalTS was run for 1 December 2022 to 28 February 2023. We refer to this new version of CalTS as CalTS-New and the old version as CalTS-Op hereafter. The two versions differ in how they perform the reliability calibration. Several changes were introduced to address known systematic biases in the forecast probabilities, particularly a pronounced overforecast bias across the tropical seas north of Australia. CalTS-New has the same forecast base and lead times as AutoFcst.

\subsection*{Thunderstorm outlooks}
The Bureau's Thunderstorm and Heavy Rainfall (TSHR) Team issues thunderstorm outlooks for stakeholders such as emergency services each morning before the DailyPoTS forecasts are issued. The lead day 1 outlook is valid for a 24 hour period from 15 UTC, starting approximately 16 hours after issue time, and having the same validity period as the lead day 1 DailyPoTS forecasts. The outlook has three categories; ``nil thunderstorm" (less than 9.5\% chance), ``thunderstorms possible" (at least 9.5\% but less than 29.5\% chance), 
and ``thunderstorms likely" (at least a 29.5\% chance)\footnote{The bins used in operations are technically $[0, 10), [10, 30), [30, 100]$. Since the forecasts are rounded to the nearest percentage point we use the bins $[0, 9.5], (9.5, 29.5], (29.5, 100]$ to align with how the elementary scoring functions are defined in Eq.~(\ref{eq:elementaryscore}).}. CalTS-Op, in conjunction with NWP models and observations are used to produce these outlooks.

\subsection*{Official forecasts}
Official forecasts are the gridded DailyPoTS forecasts that are used to generate publicly available thunderstorm forecast products on the Bureau's website and mobile phone app. Official DailyPoTS forecasts are produced in the Graphical Forecast Editor (GFE) \citep{hart2019road}, where operational meteorologists start with a copy of the AutoFcst (12 UTC run) DailyPoTS forecast grids and manually adjust them with the aim of improving the forecast service \citep{just2020streamlining}. For lead day 1 forecasts, two main adjustments are of interest for this paper:
\begin{enumerate}
\item First, DailyPoTS grids are minimally adjusted to be consistent with the Day 1 thunderstorm outlook.
\item Then, over most of the tropics in summer, whenever the daily probability of precipitation (DailyPoP) exceeds 34\%, DailyPoTS will be adjusted to be at least 30\%. This edit, whilst potentially introducing inconsistencies with the Day 1 thunderstorm outlook, will result in a thunderstorm icon on public weather forecasts.
\end{enumerate}
To remove sharp spatial artifacts introduced by such adjustments, some smoothing is applied at the final step. The Official forecast is typically issued about 8 hours before the 15 UTC start of the Day 1 validity period.

\subsection*{Observation data}
Gridded lightning observations are generated using the WeatherZone Total Lightning Network (WZTLN) data which is part of the Earth Networks Total Lightning Network \citep{rudlosky2015evaluating}. The WZTLN data is
 derived from over 100 ground-based sensors across Australia and detects both cloud-to-cloud and cloud-to-ground lightning. False positives occur less than 1\% of
 the time \citep{Price2021}. Grids of lightning counts are generated by counting lightning strikes that occurred within a 10-km radius of the centroid of a grid cell.  Strike counts were then converted to a binary lightning flag using a threshold of one strike. 

\subsection*{Processing of data}
Gridded Official, AutoFcst, and observation data were retrieved from a Bureau database with an Australian Albers equal area map projection with approximately 5.7 km grid spacing. CalTS-New and the climatological dataset were regridded to the Australian Albers equal area map projection using a bi-linear interpolation since DailyPoTS is a smooth field. Regridded forecasts were rounded back to the nearest percentage point. An equal area projection was used to ensure that results are not biased towards certain geographical areas. If any forecast system or observation has missing data for a given spatiotemporal point, that point is treated as missing data across all datasets. Additionally, matching of missing data is done across all lead days to allow comparison of performance across all seven lead days. 

\begin{figure}[h]
 \centerline{\includegraphics[width=35pc]{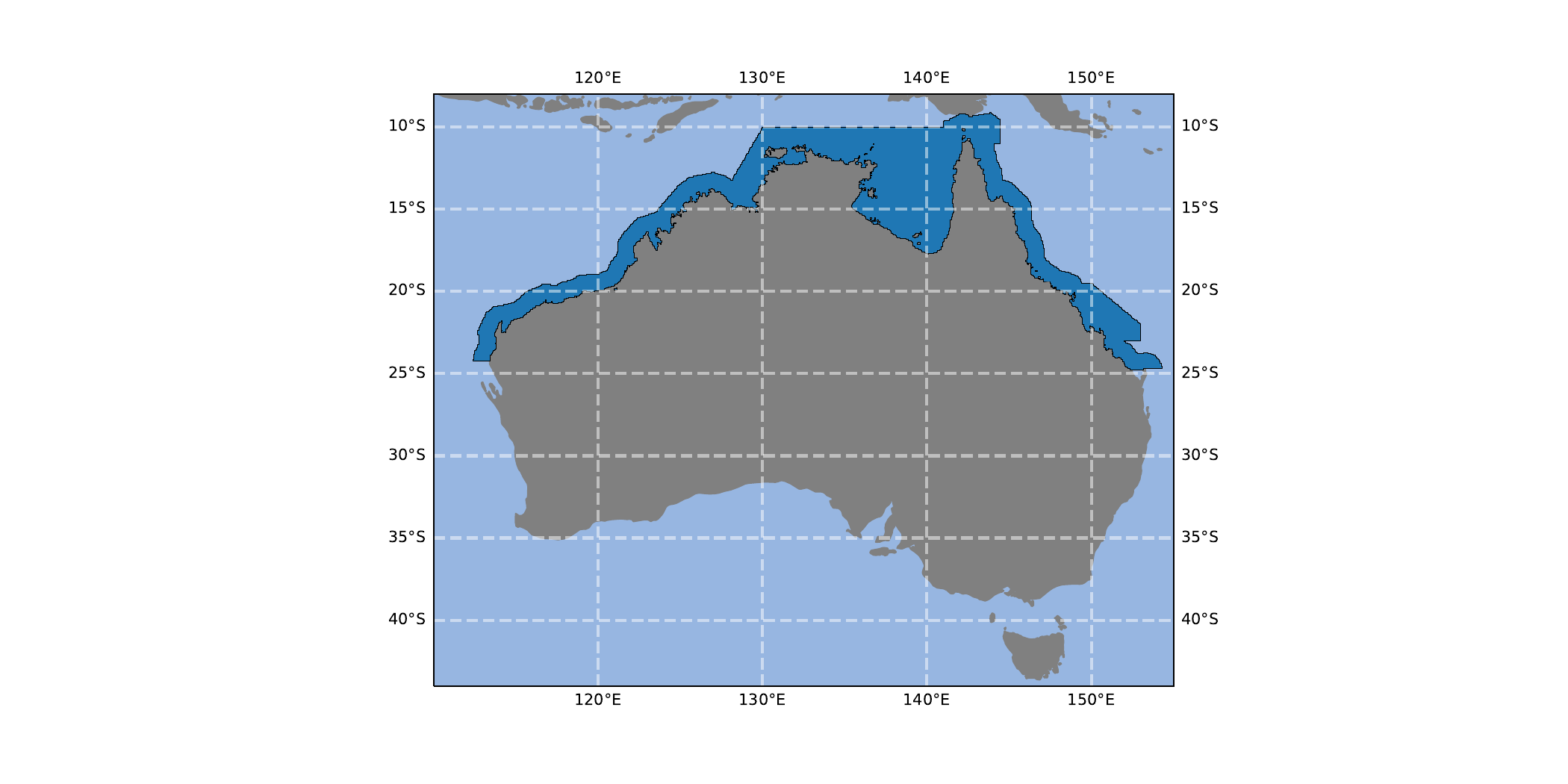}}
\caption{The tropical waters area around Australia, which lies within the domain of all forecast systems studied in the paper, are shown in dark blue to the north of Australia.}\label{fig:regions}
\end{figure}

In this paper, we demonstrate various verification methods primarily using forecasts for the Australian tropical waters (Fig.~\ref{fig:regions}). This area is of interest because CalTS-New was developed partly to address the pronounced over-forecast bias in CalTS-Op over Australian tropical waters. It is also one of the only areas where the Official DailyPoTS grids are potentially inconsistent with the Day 1 categorical thunderstorm outlook.

\section{Summarizing results}
To summarize performance for three different forecast systems across all seven lead days (1-7), we produce miscalibration--discrimination diagrams \citep{gneiting2023model, dimitriadis2023evaluating}.  These diagrams provide a useful tool for summarizing overall predictive performance for many forecast systems as well as displaying measures of calibration and discrimination in a single diagram.

\subsection*{CORP decomposition}
Miscalibration--discrimination diagrams are based on the CORP (Consistent, Optimally binned, Reproducible, and Pool-Adjacent-Violators (PAV) algorithm-based) decomposition \citep{dimitriadis2021stable}, which is given by
\begin{equation}\label{eq:mcbdsc_decomposition}
\overline{S} = \frac{1}{n}\sum_{i=1}^{n} S(x_i, y_i)  = \MCB - \DSC + \UNC,
\end{equation}
where $\overline{S}$ is the mean score for a proper scoring rule $S$ for $n$ forecast-observation pairs $(x_i, y_i)$, $\MCB$ is a measure of forecast miscalibration, $\DSC$ is a measure of forecast discrimination ability, and $\UNC$ is a measure of uncertainty which is independent of forecasts. The values of $\overline{S}$, $\MCB$, $\DSC$ and $\UNC$ are all non-negative, with lower values of $\overline{S}$ and $\MCB$ indicating better predictive performance and calibration, and higher values of $\DSC$ indicating better discrimination.

For probabilistic forecasts of binary outcomes, a classical choice of proper scoring rule $S$ is the Brier score\footnote{In this paper we refer to the ``Brier score'' as a scoring rule for a single forecast--observation pair and the ``mean Brier score'' as the mean of the Brier score across all forecast--observation pairs.}, which is defined by $S(x,y) = (x-y)^2$ for a probabilistic forecast $x$ in the unit interval $[0,1]$ and an observation $y$ taking values from the set $\{0,1\}$. In this case, the $\MCB$, $\DSC$, and $\UNC$ terms equal the reliability, resolution and uncertainty components  of the classical decomposition of the Brier score \citep{murphy1973new} if the classical bins are the set of unique forecast values with nondecreasing conditional event frequencies  \citep[][Theorem 2]{dimitriadis2021stable}.

To calculate the three terms of the CORP decomposition, we compute the mean scores $\overline{S}_\mathrm{c}$ and $\overline{S}_\mathrm{r}$ of the set of (re)calibrated forecasts $\hat x_i$ and of the best constant reference forecast $r$ (defined as the value that would yield the best score if the same value was forecast everywhere spatiotemporally), namely
\begin{equation*}
\overline{S}_\mathrm{c} = \frac{1}{n}\sum_{i=1}^{n} S(\hat{x}_i, y_i)  \quad\text{and}\quad \overline{S}_\mathrm{r} = \frac{1}{n}\sum_{i=1}^{n} S(r, y_i), \end{equation*}
where each recalibrated forecast $\hat x_i$ is obtained via isotonic regression. Isotonic regression is a method for fitting a non-decreasing free-form line to a sequence of forecast--observation pairs so that line lies as close to the observations as possible. The CORP approach uses the PAV algorithm \citep{ayer1955empirical} for the isotonic regression step on account of its computational efficiency. In this application, the CORP decomposition is a diagnostic tool and so the regression is performed on the same data that is being evaluated. This is in contrast to evaluating an isotonic regression predictive model that needs evaluating on an independent dataset. The three components are then calculated as
\begin{equation*} 
\MCB = \overline{S} - \overline{S}_c  \text{,}\quad \DSC=\overline{S}_r - \overline{S}_c\quad\text{and}\quad \UNC = \overline{S}_r.
\end{equation*}
$\MCB$ is the difference in performance based on scoring rule $S$ between the forecast and the recalibrated forecast. $\DSC$ is the difference in performance between the best constant reference forecast and the recalibrated forecast. Since the best constant reference forecast is calibrated, $\DSC$ measures the discrimination of the forecast. $\UNC$ is simply the mean score of the best constant reference forecast. $\MCB$, $\DSC$, and $\UNC$ are non-negative subject to minor conditions \citep[][Theorem 1]{dimitriadis2021stable}.

While there have been several approaches to decomposing scores into discrimination and calibration terms (e.g., \citealt{murphy1973new}), the CORP decomposition approach has several advantages. Traditional approaches often rely on making ad-hoc binning choices (e.g., choosing 10 equidistant bins). \citet{dimitriadis2021stable} showed that minor changes in binning can produce drastically different reliability diagrams. Instabilities and biases from binning choices may flow through to other miscalibration measures such as the Brier score reliability component \citep{brocker2012estimating, Ferro_2012, roelofs2022mitigating}, noting that this has a smaller impact on larger datasets. In contrast, the CORP approach optimally bins the forecasts such that no other binning scheme produces better performing recalibrated forecasts subject to monotonicity constraints \citep{brummer2013pav} and easily yields reproducible results since there is no tuning of parameters. \citet{stephenson2008two} also provided a way to account for binning problems when calculating the Brier score components by introducing two additional within-bin components. However, one significant advantage of the CORP decomposition is that it can be applied to any proper scoring rule for probabilistic forecasts of binary events and can also be extended to quantile and mean-value forecasts \citep{jordan2021characterizing}.

\subsection*{Constructing miscalibration--discrimination diagrams}
To visualize the predictive performance of the three probabilistic forecast systems using miscalibration--discrimination diagrams, we must select a proper scoring rule $S$ on which to base the CORP decomposition. We select the Brier score $S$, so that $\bar S$ is the mean Brier score. The implications of this choice, and of other possible choices, will be explored in sections~\ref{triptych} and~\ref{mcb-dsc_FIRM}.

The miscalibration--discrimination diagram is a scatter plot of $\DSC$ against $\MCB$ with parallel, diagonal lines showing the mean score $\overline{S}$. $\UNC$ is also plotted as a diagonal line that intersects with the origin, since $\MCB = \DSC = 0$ for the best constant reference forecast. These diagrams provide a useful tool for easily comparing overall predictive performance, discrimination, and calibration for many forecast systems and can be applied to whatever proper scoring rule is required. They are interpreted in the following way:
\begin{enumerate}
\item Points on a higher diagonal rung (i.e. towards the top left) have better overall predictive performance.
\item Points higher up vertically have higher discrimination, or equivalently the highest potential score subject to recalibration.
\item Points further to the left have better calibration.
\end{enumerate}

We make several small changes to the original miscalibration--discrimination diagrams \citep{gneiting2023model, dimitriadis2023evaluating}, as illustrated in Fig.~\ref{fig:brier_mcb_dsc}. 
\begin{enumerate}
\item We are using gridded rather than single point forecasts and observations. Our approach here is to perform isotonic regression and calculation of the best constant reference forecast across the region rather than individually at every grid point. Accompanied by corresponding reliability diagrams (Section~\ref{triptych}), this has the advantage of providing actionable information for meteorologists to inform suitable conditional bias correction strategies over a geographical area. 
\item We add a diagonal reference line based on a long term climatological forecast for each grid point and each day of the year. Unlike the best constant reference forecast, the values in this reference forecast vary spatiotemporally. This provides a practical benchmark that is an alternative to the best constant reference forecast, since it is known \textit{a priori}. See Appendix~A for details of its construction.
\item We plot forecast performance across multiple lead days and connect the points for a given forecast system by a line. 
\item We generate 95\% confidence intervals via circular block bootstrapping \citep{wilks1997resampling}. Blocks are taken across dimensions $x$, $y$, and $time$ with block length being the square root of the length of the dimension. Confidence intervals are calculated for both the $\MCB$ and $\DSC$ components so that they appear as a cross on the diagram. 
\end{enumerate}

\subsection*{Miscalibration--discrimination diagram results}
\begin{figure}[h]
 \centerline{\includegraphics[width=42pc]{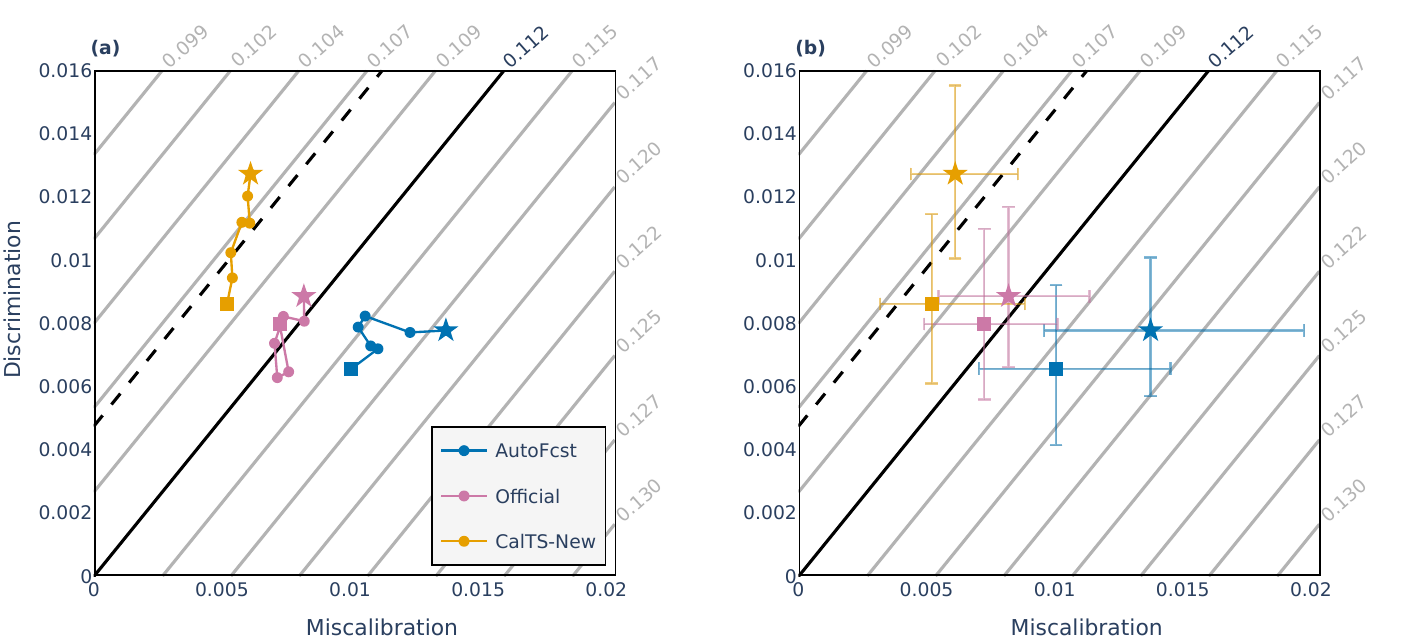}}
  \caption{Miscalibration--discrimination summary scores based on the Brier score. Scores of MCB against DSC are plotted for each lead day, with the gray diagonal lines indicating the mean Brier score. MCB and DSC are analogous to the reliability and resolution components of the classical Brier score decomposition.  Stars indicate a lead day 1 forecast while squares indicates a lead day 7 forecast. The solid black line corresponds to the score of the best constant reference forecast. The dashed black line displays the score of the long-term spatiotemporally varying climatological reference forecast. (a) Scores are shown for all seven lead days. (b) Confidence intervals (generated through circular block bootstrapping with 1000 resamples) are shown for lead day 1 and 7 forecasts.
}\label{fig:brier_mcb_dsc}
\end{figure}

Miscalibration--discrimination diagrams based on the Brier score for our three forecast systems are shown in Fig.~\ref{fig:brier_mcb_dsc}. Confidence intervals for the MCB and DSC components are shown for lead day 1 and lead day 7 forecasts in Fig.~\ref{fig:brier_mcb_dsc}b. If confidence intervals are added for all seven lead days, the figure becomes too crowded. However, if confidence intervals for all lead days or more forecast systems are required on the miscalibration--discrimination diagram, then we recommend producing interactive plots that allow one to toggle lines on and off rather than static figures.

The miscalibration--discrimination diagram highlights that AutoFcst had no skill at any lead day based on the mean Brier score compared to the best constant reference forecast for the region. Official forecasts had better predictive performance than AutoFcst but performed similarly to the constant reference forecast. The gain in predictive performance was primarily due to meteorologists improving calibration rather than discrimination. CalTS-New showed superior performance based on the mean Brier score, outperforming the best constant reference forecast across all lead days, and was slightly better than the long-term climatological forecast at shorter lead days. This was due to improvements in both discrimination and calibration. For a well calibrated forecast system, it is expected that the improvement in skill from longer to shorter lead times should come from better discrimination, while calibration remains fairly constant. Note that the confidence intervals are for the MCB and DSC terms and that if one wants to calculate the likelihood that a forecast performed better than a reference forecast or another forecast sources, then those confidence intervals should be calculated using $\bar S$. Miscalibration--discrimination diagrams were also produced for eight other geographical areas across Australia (not shown) and in these cases all three forecast systems performed better than both the best constant reference forecast and the long-term climatological forecast.

%%%%%%%%%%%%%%%%%%%%%%%%%%%%%%%%%%%%%
\section{Limitations of categorical performance diagrams}\label{s:performance diagrams}

In the previous section we assessed overall predictive performance of probabilistic forecasts for thunderstorms using a proper scoring rule (the Brier score), with results presented on miscalibration--discrimination diagrams. Another approach to assessing overall predictive performance, alongside other characteristics like discrimination, that has become increasingly popular within sections of the meteorological research community is to use categorical performance diagrams. Categorical performance diagrams summarize multiple verification measures for binary forecasts in a single diagram \citep{roebber2009visualizing}.
While initially created to evaluate dichotomous forecasts and warnings, a number of recent studies have used these diagrams to verify probabilistic forecasts of thunderstorms \citep{harrison2022utilizing}, severe convective weather \citep{loken2017comparison, gagne2017storm,  cintineo2020noaa, lagerquist2020deep, Flora_2021, gallo2022exploring, miller2022exploring, sandmael2023tornado} or snow \citep{uden2023evaluation, radford2023improving}. While, performance diagrams may be suitable for measuring discrimination ability, in this section, we explain why some interpretations of categorical performance diagrams lead to misguided inferences regarding overall predictive performance.

\begin{figure}[h]
 \centerline{\includegraphics[width=18pc]{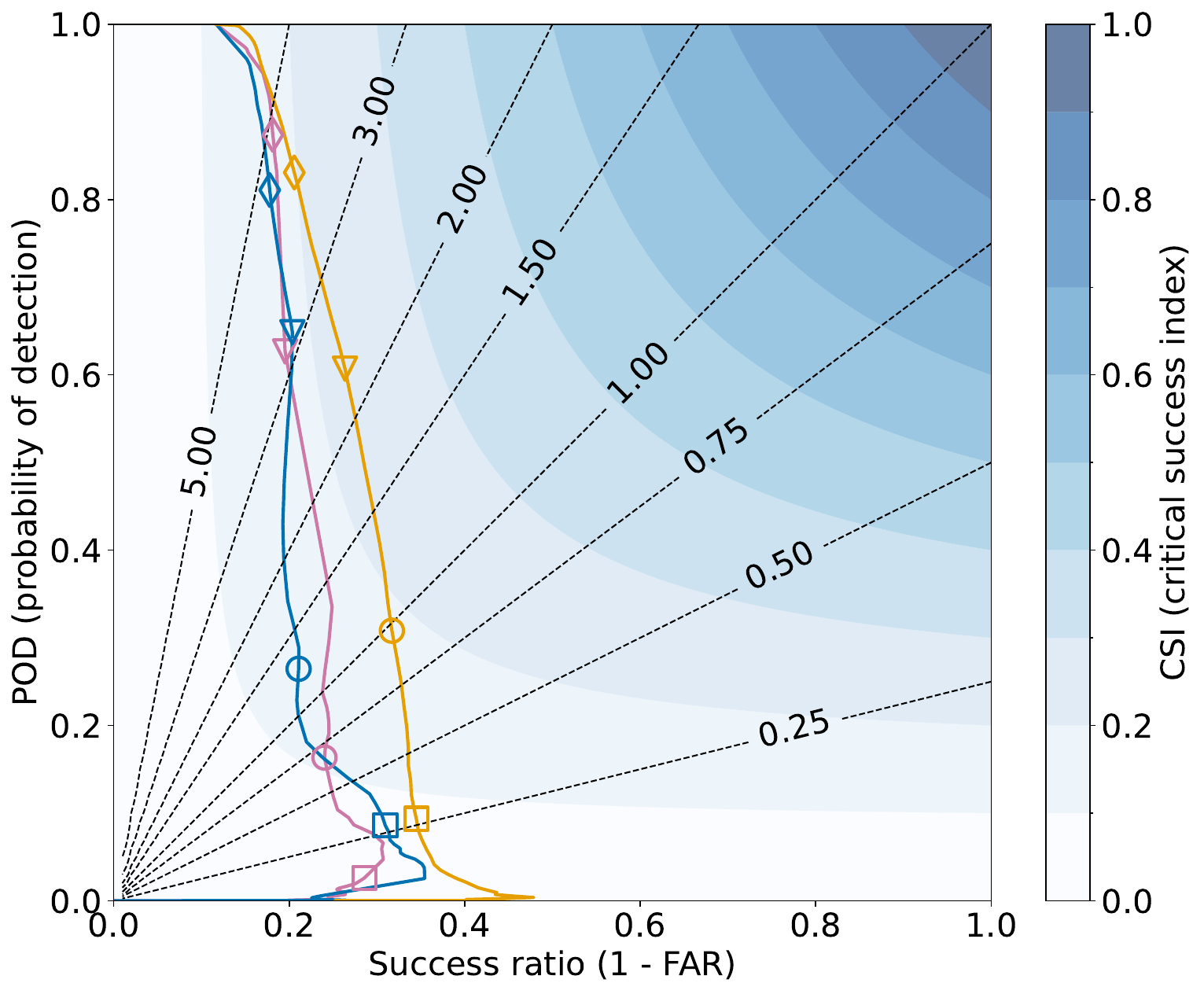}}
  \caption{Performance diagram for probability of thunderstorm forecasts issued by AutoFcst (blue), Official (pink) and CalTS-New (orange). The markers correspond to threshold probabilities of 20\% (diamonds), 30\% (triangles), 40\% (circles) and 50\% (squares).
}\label{fig:performance diagram ts data}
\end{figure}

Figure~\ref{fig:performance diagram ts data} illustrates a performance diagram for the set of probability of thunderstorm forecasts issued by AutoFcst, Official and CalTS-New. The diagram shows values of POD on the vertical axis, SR on the horizontal axis, FB via dotted diagonal lines and CSI, via blue contoured shading. These are verification measures for a set of binary forecasts, defined by the formulae
\[\mathrm{POD} = \frac{h}{h + m}, \quad \mathrm{SR} = \frac{h}{h+f}, \quad \mathrm{FB} = \frac{h+f}{h+m} \quad\mbox{and}\quad \mathrm{CSI} = \frac{h}{h+m+f},\]
where $h$, $m$ and $f$ are the number of hits, misses and false alarms for the set of forecast cases. To calculate POD and SR for a set of probabilistic forecasts, a threshold probability $\theta$ is selected and each probabilistic forecast $p$ is converted into a forecast in binary space of `event' if $p \geq \theta$ and `non-event' if $p < \theta$. For example, the circle markers on Figure~\ref{fig:performance diagram ts data} show POD and SR values when the threshold probability 40\% is selected. By plotting points for many different threshold probabilities, performance diagram curves are generated. By swapping the axes of the diagram, so that SR (sometimes called \textit{precision}) is plotted against POD (sometimes called \textit{recall}), one obtains \textit{precision--recall curves} \citep{manning1999foundations, davis2006relationship}. The area under a precision--recall curve (AUCPR) equals the area to the left of the performance diagram curve.

Several authors have recently used performance diagrams to rank the overall predictive performance of forecast systems issuing probability forecasts. Most commonly, the forecast system with the highest maximum CSI value is judged to have performed best compared to its competitors.\footnote{The authors also found an instance in the meteorological literature where the system with the highest mean CSI was judged to be superior. By adapting arguments shown here, it can be shown that using mean CSI to rank overall predictive performance can lead to misguided inferences.} Another tempting approach may be to use the area to the left of the performance diagram curve for ranking.\footnote{In the meteorological literature, this area is sometimes imprecisely referred to as the area under the curve.} While both approaches may have uses for ranking discrimination ability, they lead to unsound inference if used to rank overall predictive performance. The use of CSI for this purpose is problematic because it is not a proper scoring rule for probabilistic forecasts of binary events. In Appendix~B we show that maximizing CSI can result in issuing forecasts that go against the forecaster's best judgment. As discussed in Section~\ref{triptych}, maximum CSI can be interpreted as a simple measure of discrimination whilst AUCPR may be suitable for ranking \textit{discrimination} ability.

 For the remainder of this section, we construct a simple experiment that illustrates the failure of maximum CSI and AUCPR to correctly rank overall predictive performance of competing forecast systems.

Consider a probability distribution supported on the interval $[0,0.5]$. Probabilities $p_i$ are drawn from this distribution, so that $0\leq p_i\leq 0.5$,  and for the $i$th case the realization (event or no event) is drawn at random with probability of an event equal to $p_i$. We consider three forecast systems. The \textit{Ideal} forecast issues the forecasts $p_i$, and makes optimal use of the information known from draws from the first distribution. It is reliable. The \textit{Under} forecast issues the forecasts $p_i/2$, while the \textit{Over} forecast issues the forecasts $2p_i$. Both of these synthetic forecasts have the same discrimination ability as Ideal, but neither is reliable because they either under-predict or over-predict probabilities. By construction, Ideal has best overall predictive performance of the three systems.

A little thought shows that the performance diagram curves for systems Ideal, Under and Over coincide, and their maximum CSI values are identical. This is because for any threshold probability $\theta$ in the interval $[0,0.5]$, the corresponding POD, SR and CSI values for the Ideal forecast will equal the corresponding POD, SR and CSI values for the Under (respectively Over) forecast at threshold probability $\theta/2$ (respectively $2\theta$). So maximum CSI and the AUCPR ranks these three systems equally, whereas by construction the Ideal forecast is clearly superior.

\begin{figure}[h]
 \centerline{\includegraphics[width=18pc]{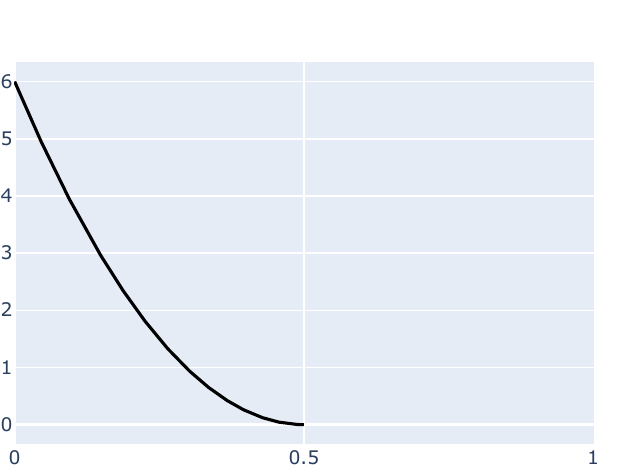}}
  \caption{The probability density function of $\mathrm{Beta}(1, 3, [0, 0.5])$.
}\label{fig:beta}
\end{figure}

\begin{figure}[h]
 \centerline{\includegraphics[width=22pc]{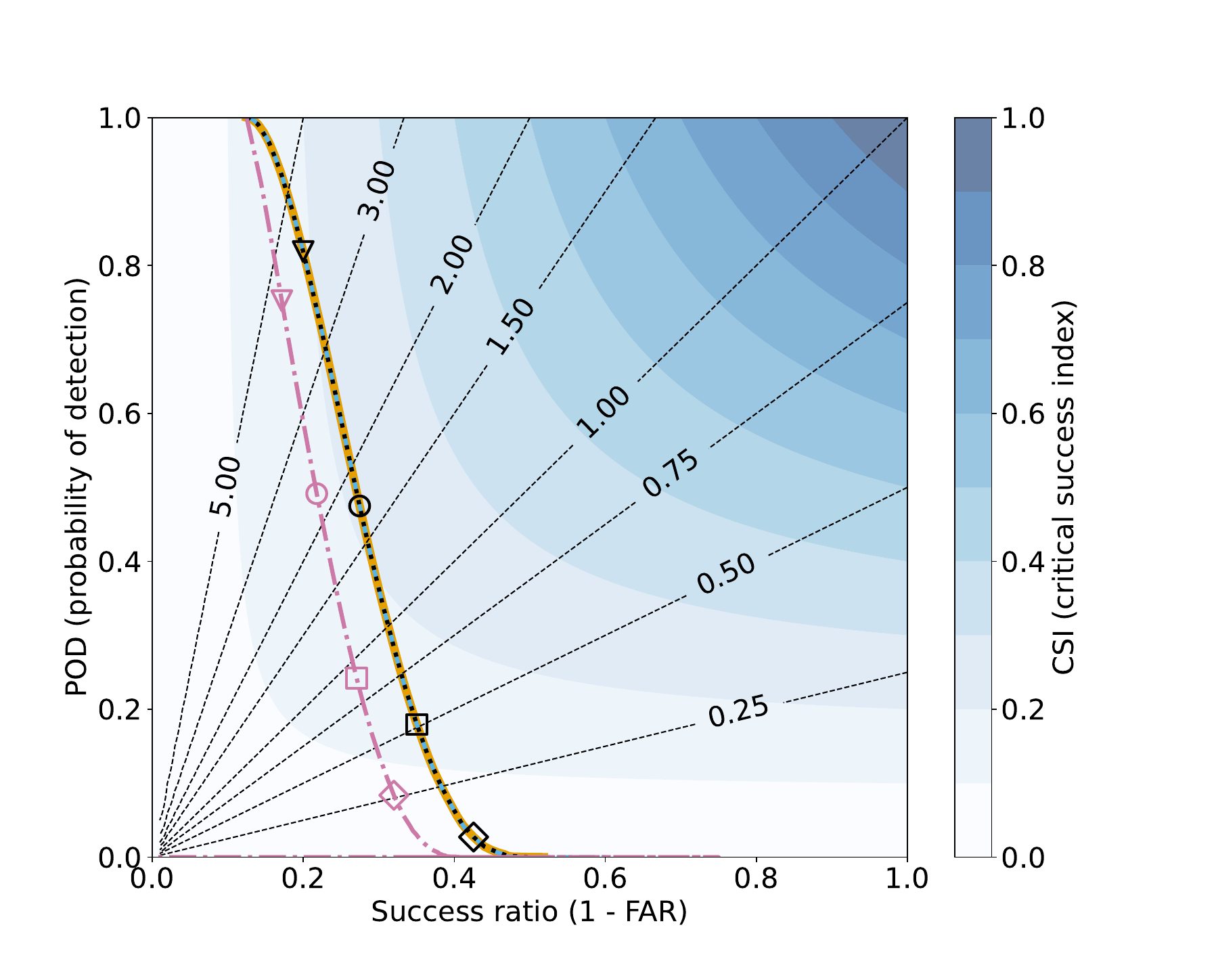}}
  \caption{Performance diagram for the synthetic experiment. Curves for the Ideal (solid orange), Under (dotted black) and Over (dashed blue) forecast systems coincide, while the Jitter (dot--dash pink) curve stands apart. The markers indicate points on the Ideal and Jitter curves using threshold probabilities of 10\% (triangle), 20\% (circle), 30\% (square) and 40\% (diamond). These correspond to threshold probabilities of 5\%, 10\%, 15\% and 20\% for Under and 20\%, 40\%, 60\% and 80\% for Over.
}\label{fig:synthetic pd}
\end{figure}

\begin{figure}[h]
 \centerline{\includegraphics[width=18pc]{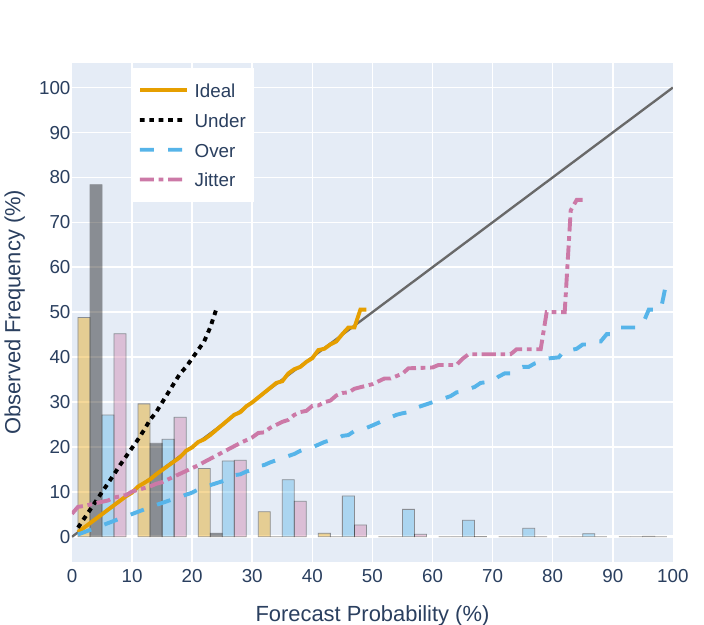}}
  \caption{Reliability curves for the synthetic experiment. Histograms indicate frequency of forecasts for each bin of width 10\%.
}\label{fig:synthetic rd}
\end{figure}

We illustrate this thought experiment using synthetic data so that probabilities $p_i$ are sampled from the beta distribution $\mathrm{Beta}(1, 3, [0, 0.5])$ with shape parameters 1 and 3 supported on the interval $[0, 0.5]$. Its probability density function is illustrated in Fig.~\ref{fig:beta}, and the relative frequency of an event with this distribution is $1/8$. For illustrative purposes, we also introduce a forth forecast system \textit{Jitter}, which issues the forecasts $p_i + r_i$, clipped to the interval $[0,1]$, where $r_i$ is random noise selected at random from the normal distribution with mean 0 and standard deviation 0.1. Jitter should therefore be reasonably reliable, but have poorer discrimination than Ideal, Under and Over. Ten million trials were run (i.e., $1\leq i \leq 10^7$) from which performance and reliability diagrams were generated (Figures~\ref{fig:synthetic pd} and \ref{fig:synthetic rd}). We also ran the experiment 1000 times, each with $10^5$ trials, and calculated statistics for each experiment. The mean of those statistics across all experiments is presented in Table~\ref{tab:synthetic results}.

\begin{table}[t]
\caption{Maximum CSI, mean Brier score and AUCPR for the synthetic experiment, with standard errors in parentheses.
}\label{tab:synthetic results}
\begin{center}
\begin{tabular}{| l | r | r | r |} \hline
Forecast system	& Max CSI					& AUCPR					& Mean Brier score \\ \hline
Ideal			& 0.214 ($7.0\times10^{-5}$)	& 0.275 ($1.1\times10^{-4}$) 	& 0.100 ($2.2\times10^{-5}$)\\
Under		& 0.214 ($7.0\times10^{-5}$)	& 0.275 ($1.1\times10^{-4}$)	& 0.106 ($2.7\times10^{-5}$)\\
Over			& 0.214 ($7.0\times10^{-5}$)	& 0.275 ($1.1\times10^{-4}$)	& 0.125 ($1.8\times10^{-5}$)\\
Jitter			& 0.178 ($6.6\times10^{-5}$)	& 0.224 ($9.8\times10^{-5}$)	& 0.108 ($2.2\times10^{-5}$)\\
\hline
\end{tabular}
\end{center}
\end{table}

As anticipated above, the performance diagram fails to distinguish between the Ideal, Under and Over forecast systems, while the Jitter system has the lowest maximum CSI and AUCPR of any forecast system. The reliability diagram highlights that Ideal is well-calibrated, while the other systems have calibration issues. The combination of both diagrams would probably indicate that Ideal has best predictive performance. However, it does not seem obvious to the authors how to rank the remaining systems. For example, Jitter mostly has better reliability properties than Over, but has poorer discrimination as measured by performance diagram indicators. If ranked for overall predictive accuracy by the Brier score, Ideal performs best followed by Under, Jitter and Over. By looking at the difference in mean Brier scores, these rankings were found to be statistically significant at the 2.5\% level.\footnote{For example, the difference between the mean Brier scores for Jitter and Under was positive in more than 97.5\% of experiments.}

This synthetic experiment was repeated using a range of different shape parameters for the beta distribution and the overall story was similar: Ideal, Over and Under had the same maximum CSI and AUCPR while Jitter had lower such values, and the mean Brier score correctly identified Ideal as the most accurate forecast overall.

%%%%%%%%%%%%%%%%%%%%%%%%%%%%%%%%%%%%%%%%%
\section{The triptych approach for verification of probability forecasts of binary outcomes} \label{triptych}
Categorical performance diagrams were designed to convey multiple verification measures on one diagram for assessment and diagnostic purposes and can be used to measure discrimination ability. A more comprehensive approach for forecast evaluation, is the triptych approach of \citet{dimitriadis2023evaluating}. With three diagrams, the reliability, discrimination ability, and overall predictive performance and value is diagnosed for each forecast system.

%%%%%%%%%%%%%%%
\subsection*{Murphy Diagrams}
Murphy diagrams \citep{hernandez2011brier, ehm2016quantiles} are a sound diagnostic for visualizing predictive performance of probability forecasts as a function of user decision threshold since, unlike the CSI used in performance diagrams, they are based on proper scoring rules. In the first half of this subsection, the theory that justifies the use of Murphy diagrams will be introduced. The second half will provide illustrative examples from Bureau forecast systems and the synthetic experiment of Section~\ref{s:performance diagrams}.

\citet{schervish1989general} and \citet{ehm2016quantiles} showed that, subject to technical conditions, each proper scoring rule $S$ admits a representation\footnote{The representation of Eq.~(\ref{eq:mixture representation}) is known as as a \textit{mixture representation}, and the measure $H$ as the \textit{mixing measure}.} of the form
\begin{equation}\label{eq:mixture representation}
S(x, y) = \int_{0}^{1} S_\theta(x, y) \, \dd H(\theta),
\end{equation}
where $H$ is a non-negative measure and for each $\theta$ in $(0,1)$ the scoring function $S_\theta$ is given by
\begin{equation} \label{eq:elementaryscore}
 S_{\theta}(x, y) =  
	\begin{cases}
           2\theta, & \text{if } y=0,\, x > \theta, \\
           2(1-\theta), &  \text{if } y=1,\, x\leq\theta, \\
           0, & \text{otherwise.}
	\end{cases}
\end{equation}
In the following, we explain that the parameter $\theta$ can be interpreted as a probabilistic decision threshold, the score $S_\theta(x,y)$ as proportional to the economic loss for a user with that threshold, and the measure $H$ as a weighting to apply across all decision thresholds to recover the overall score $S(x,y)$. In a nutshell, Eq.~(\ref{eq:mixture representation}) says that each proper scoring rule $S$ can be written as a weighted average of economic regret across all user decision thresholds.

The scoring function $S_{\theta}$ essentially assigns a penalty, equal to twice the distance between the decision threshold $\theta$ and the observation $y$, whenever the forecast $x$ and observation $y$ lie on different sides of the decision threshold $\theta$. Otherwise no penalty is given. As shown by \citet{ehm2016quantiles}, the score $S_\theta(x,y)$ is proportional to the economic loss of a user with decision threshold $\theta$ who takes action based on the forecast $x$ within the classical cost--loss decision model with cost--loss ratio $\theta$ \citep[c.f.][]{richardson2000skill}. Importantly, $S_\theta$ is a \textit{proper} scoring rule \citep{ehm2016quantiles}. Since Eq.~(\ref{eq:mixture representation}) shows that each proper scoring rule $S$ can be represented as an integral of scoring functions $S_\theta$, each $S_\theta$ is called an \textit{elementary scoring function}. Moreover, Eq.~(\ref{eq:mixture representation}) has the following interpretation: the score $S(x,y)$ is the economic regret averaged over all user decision thresholds $\theta$, weighted by the measure $H(\theta)$.

By selecting an appropriate measure, Eq.~(\ref{eq:mixture representation}) can be used to find a proper scoring rule that weights each decision threshold as desired. For example, if one desires to weight each decision threshold equally, then the uniform measure $\dd H(\theta) = \dd\theta$ can be selected. A simple calculation, shown in Appendix~C, shows that the corresponding proper scoring rule $S$ is the Brier score. Hence the Brier score can be interpreted as a proper scoring rule that weights all decision thresholds equally. On the other hand, if one desires to put a very high weight on decision thresholds near 0 and 1 whilst de-emphasizing middling probabilistic decision thresholds, then the measure $\dd H(\theta) = (2\theta(1-\theta))^{-1}\,\dd\theta$ could be chosen. A similar calculation shows that this choice obtains the logarithmic score.

\begin{figure}[h]
 \centerline{\includegraphics[width=22pc]{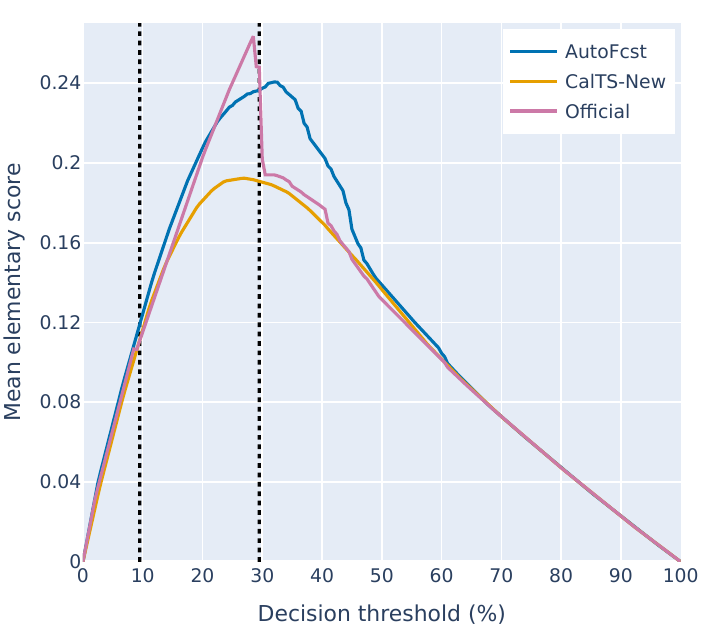}}
  \caption{Murphy diagrams for daily probability of thunderstorm forecasts for Official, AutoFcst, and CalTS-New lead day 1 forecasts. Lower mean elementary scores are better. Probabilistic decision thresholds 9.5\% and 29.5\% are highlighted with vertical dotted lines.
}\label{fig:murphy}
\end{figure}

\begin{figure}[h]
 \centerline{\includegraphics[width=18pc]{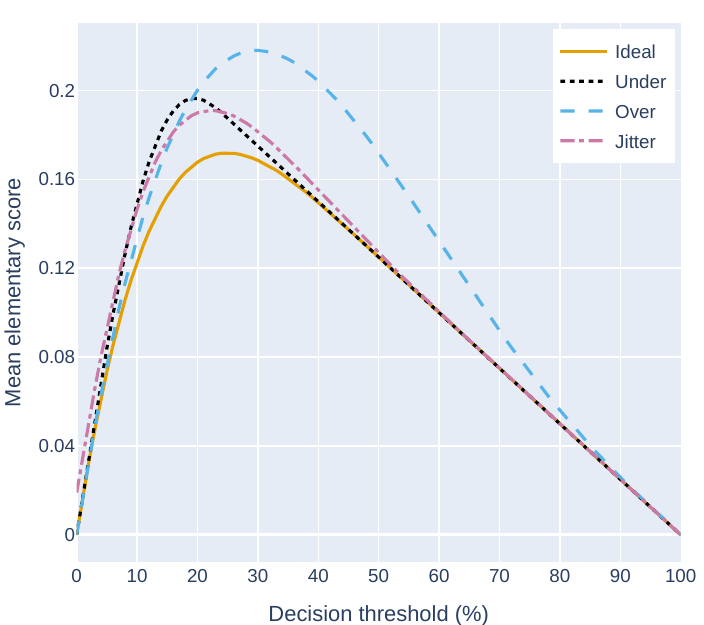}}
  \caption{Murphy diagram for the synthetic experiment.
}\label{fig:murphy synthetic}
\end{figure}

The Murphy diagram \citep{ehm2016quantiles} for a set of $n$ forecast--observation pairs $(x_i,y_i)$ is a plot of the mean elementary scores $\bar{S}_\theta$ against decision threshold $\theta$, where
\[\bar{S}_\theta=\frac{1}{n}\sum_{i=1}^n S_\theta(x_i,y_i).\]
See Figures~\ref{fig:murphy} and \ref{fig:murphy synthetic} for examples. A lower curve is better. The area under the plotted curve is the mean Brier score, which follows by taking means across the $n$ forecast cases in Eq.~(\ref{eq:mixture representation}) with uniform measure $H$. If the mean elementary scores for two forecast systems are plotted on the same diagram, and if one curve always lies beneath the other, then it follows from Eq.~(\ref{eq:mixture representation}) that the forecast system with the lower curve outperforms the other system for \textit{every} proper scoring rule $S$ \citep{ehm2016quantiles}. On the other hand, if the curves cross, then it is possible to construct two proper scoring rules that rank their performance differently \citep[c.f.][]{merkle2013choosing} via judicious choice of measures $H$.

Fig.~\ref{fig:murphy} shows Murphy diagrams for lead day 1 thunderstorm forecasts. Interestingly, the difference in performance between Official and AutoFcst varies greatly across some user decision thresholds. For user decision thresholds between 22\% and 29\%, Official forecasts performed worse than AutoFcst, while it matched or bettered AutoFcst for other decision thresholds. Since we have this interesting variation in performance across decision thresholds, one can easily see how different scoring rules may rank the performance of Official and AutoFcst differently depending on the measure $H$ in Eq.~(\ref{eq:mixture representation}).  The spike in mean elementary score for Official forecasts occurs at the 28\% decision threshold and occurs due to meteorologists regularly clipping maximum probability of thunderstorm forecasts to be less than 30\%.  This has the impact of improving the forecast performance for users who care about decision thresholds of at least 30\%, but degrading the forecast performance for users who make decisions for thresholds just below 30\%.

Figure~\ref{fig:murphy synthetic} shows a Murphy diagram for the synthetic experiment of Section~\ref{s:performance diagrams}. Note that the Murphy curve for Ideal (orange) lies beneath the other Murphy curves for decision thresholds up to 50\%, and lies on or below the other curves for higher decision thresholds. Hence Ideal will correctly be identified as the superior forecasting system over this set of forecast cases when measured by any \textit{strictly} proper scoring rule (i.e., a proper scoring rule, like the Brier score, where the measure $H$ places positive weight on every decision threshold). This is consistent with the fact that, by construction, Ideal issues forecasts that makes optimal use of the available information from the beta distribution, whilst the other systems either distort that information or add noise (c.f. \citet{gneiting2011comparing}). The predictive dominance of Ideal cannot be deduced from the performance diagram of Fig.~\ref{fig:synthetic pd}, nor from some proper scoring rules that are not strictly proper. For example, the Murphy curves for Ideal and Under coincide for decision thresholds exceeding 50\% because neither forecast system predicts probabilities exceeding 0.5, and so any proper score that places zero weight on decision thresholds up to 50\% will rank Ideal and Under equally. If predictive performance around the 20\% threshold were targeted by a proper scoring rule, Jitter would rank better than Under and Over.

%%%%%%%%%%%%%%%%%%%
\subsection*{Murphy Diagram differences}
\begin{figure}[h]
 \centerline{\includegraphics[width=37pc]{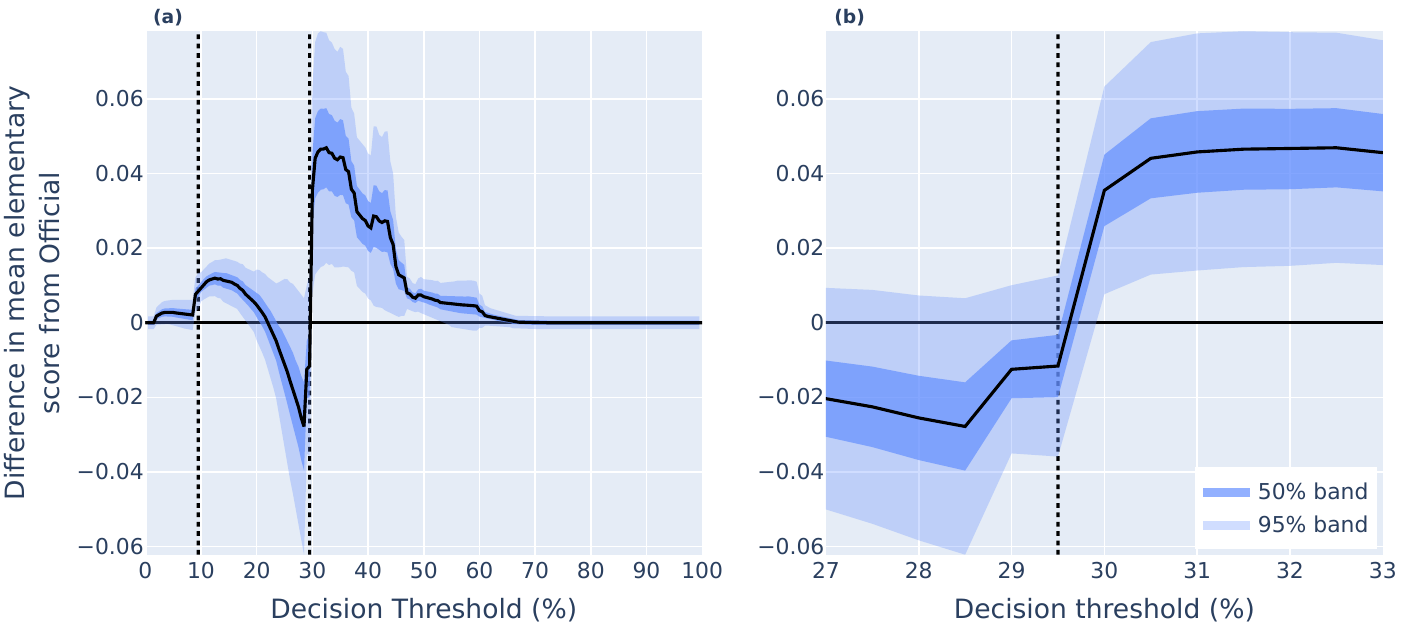}}
  \caption{The difference in mean elementary score between AutoFcst and Official. Positive values indicate that AutoFcst had a higher (worse) mean elementary score for a given decision threshold than Official. Dark and light blue shading shows 50\% and 95\% confidence bands calculated using the Diebold--Mariano test statistic. Probabilistic decision thresholds 9.5\% and 29.5\% are highlighted with vertical dotted lines. (a) shows the difference across all thresholds. (b) only shows decision thresholds ranging from 27\% to 33\%.
}\label{fig:murphy_ci}
\end{figure}

To understand the difference in performance between two forecast systems, we take the difference in elementary scores between forecast systems and calculate confidence intervals using the \citet{hering2011comparing} modification of the Diebold--Mariano test statistic \citep{diebold1995comparing} (Fig.~\ref{fig:murphy_ci}). The Diebold--Mariano test statistic has advantages over many other methods for comparing two forecasts as it accounts for both serial and contemporaneous temporal correlation, and forecast errors can be non-Gaussian. Spatial means of the difference in the elementary scores are taken before calculating the test statistic to account for any spatial correlation.

The difference in Murphy curves clearly highlights the difference in performance between Official and AutoFcst for each decision threshold. Meteorologists focus closely on thresholds 9.5\% and 29.5\% and the improvements that they made over AutoFcst at the 9.5\% decision threshold were statistically significant at the 2.5\% level. The differences are not statistically significant for the 29.5\% decision threshold, but are for decision thresholds between 30\% to 53\%. This is likely because clipping the maximum DailyPoTS value to be 29\% when ``possible''  is forecast on the thunderstorm outlooks improves the forecast, while increasing the DailyPoTS to 30\% when DailyPoP exceeds 34\% acts against these improvements at the 29.5\% decision threshold.

%%%%%%%%%%%%%%%%%%
\subsection*{CORP reliability diagrams}
\begin{figure}[h]
 \centerline{\includegraphics[width=37pc]{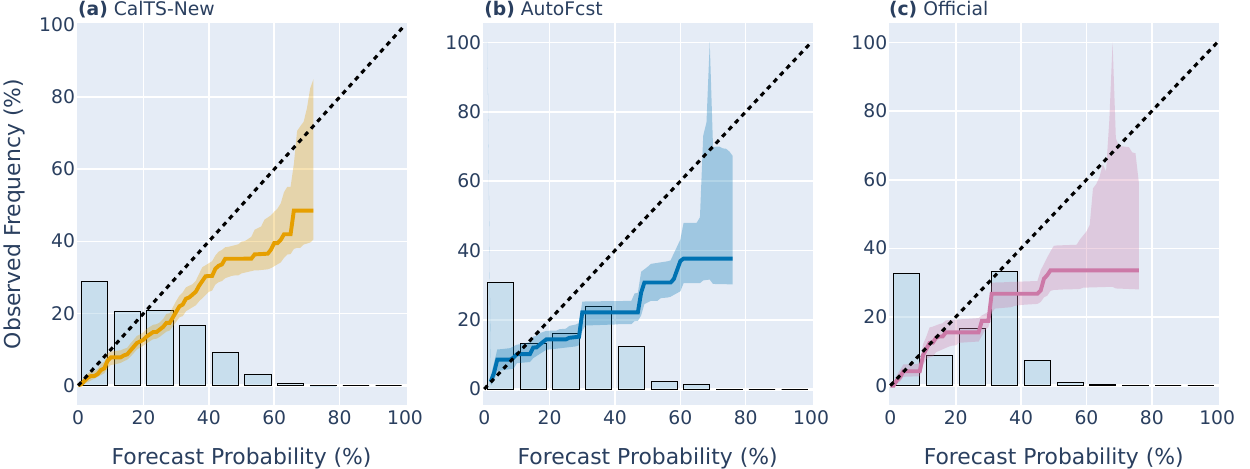}}
  \caption{CORP Reliability diagrams for lead day 1 forecasts for (a) CalTS-New, (b) AutoFcst, and (c) Official with 95\% confidence bands. A histogram indicates the frequency that forecasts were issued for probabilities located within bins that are ten percentage points wide. Bins are right open (i.e., [a, b) ), except for the final bin which includes 100\%.
}\label{fig:reliabilitywatertropics}
\end{figure}

Reliability diagrams, sometimes presented in the form of an attributes diagram \citep{hsu1986attributes}, are regularly used to visualize calibration across threshold probabilities. As noted above, \citet{dimitriadis2021stable} illustrated that traditional approaches to creating reliability diagrams are sensitive to the choice of bins and that the use of the CORP (isotonic regression) approach addresses this binning issue. We produce CORP reliability diagrams for lead day 1 forecasts  with 95\% confidence bands generated using circular block bootstrapping from 1000 resamples (Fig.~\ref{fig:reliabilitywatertropics}). Alternatively, consistency bands could be produced \citep[e.g.,][]{Br_cker_2007, dimitriadis2021stable}. A histogram of forecast frequency using 10 equidistant bins is displayed on the reliability diagrams. Note that these forecast frequency bins are not the same as the bins used to create the reliability curves.

Fig.~\ref{fig:reliabilitywatertropics} shows that all forecast systems have a general overforecast bias across most thresholds. However, CalTS-New has a smaller overforecast bias than AutoFcst.
When comparing the histograms between Official and AutoFcst, the [30\%, 40\%) bin count is much higher for Official than AutoFcst. This is due to operational meteorologists increasing the DailyPoTS value to exactly 30\% over most of the water tropics area when both the DailyPoP exceeds 34\% and the thunderstorm outlook category is ``possible''. Operational meteorologists also reduced the over-forecast bias of AutoFcst, for probabilities above 30\% and below 60\%. This was likely due to DailyPoTS being clipped to have a maximum value of either 29 or 30\% whenever ``possible'' was the forecast category on the thunderstorm outlooks. These results highlight that there is the opportunity for meteorologists to update their standard ``grid editing" procedures and for forecast system developers to continue to work to improve calibration. A user making decisions on these forecasts could use these diagrams. For example, a user with a decision threshold of 40\% and with access to AutoFcst forecasts would take action if the forecast value was around 23\%.

Finally, if inspecting reliability diagrams (Fig.~\ref{fig:reliabilitywatertropics}) with the performance diagrams (Fig.~\ref{fig:performance diagram ts data}) in isolation, it is unclear which forecast system ranks best for overall predictive performance out of AutoFcst and Official, which further demonstrates the need to measure overall predictive performance with Murphy Diagrams or an appropriate proper scoring rule.

%%%%%%%%%%%%%%%%%%%
\subsection*{Understanding discrimination ability}
While understanding the reliability or calibration of the forecasts is important, it is also worth investigating discrimination ability. This is because even perfectly calibrated forecasts, such as the best constant value forecast, may show little or no ability to discriminate between events and non-events. The $\DSC$ component of the CORP decomposition is a measure of discrimination ability. Performance diagrams (or precision--recall curves) are a diagnostic tool from which discrimination measures can be derived. Another diagnostic for discrimination ability is the receiver operating characteristic (ROC) curve \citep{mason1982model}, which has been commonly displayed alongside reliability diagrams. Like performance diagram curves, ROC curves are produced by converting each probabilistic forecast into a binary forecast using a set increasing threshold probabilities. For each threshold probability a point on the ROC curve is generated by plotting the hit rate (probability of detection) $h / (h +m)$ against the false alarm rate (probability of false detection) $f / (c + f)$, where $h$, $m$, $f$, and $c$ denote the number of hits, misses, false alarms and correct negatives from the set of binary forecasts. ROC curves can be used to understand how discrimination ability varies by threshold probability, and reflect the ``potential predictive ability'' of the forecasts if they were to be perfectly re-calibrated \citep{wilks2011statistical}. 

\begin{figure}[h]
 \centerline{\includegraphics[width=37pc]{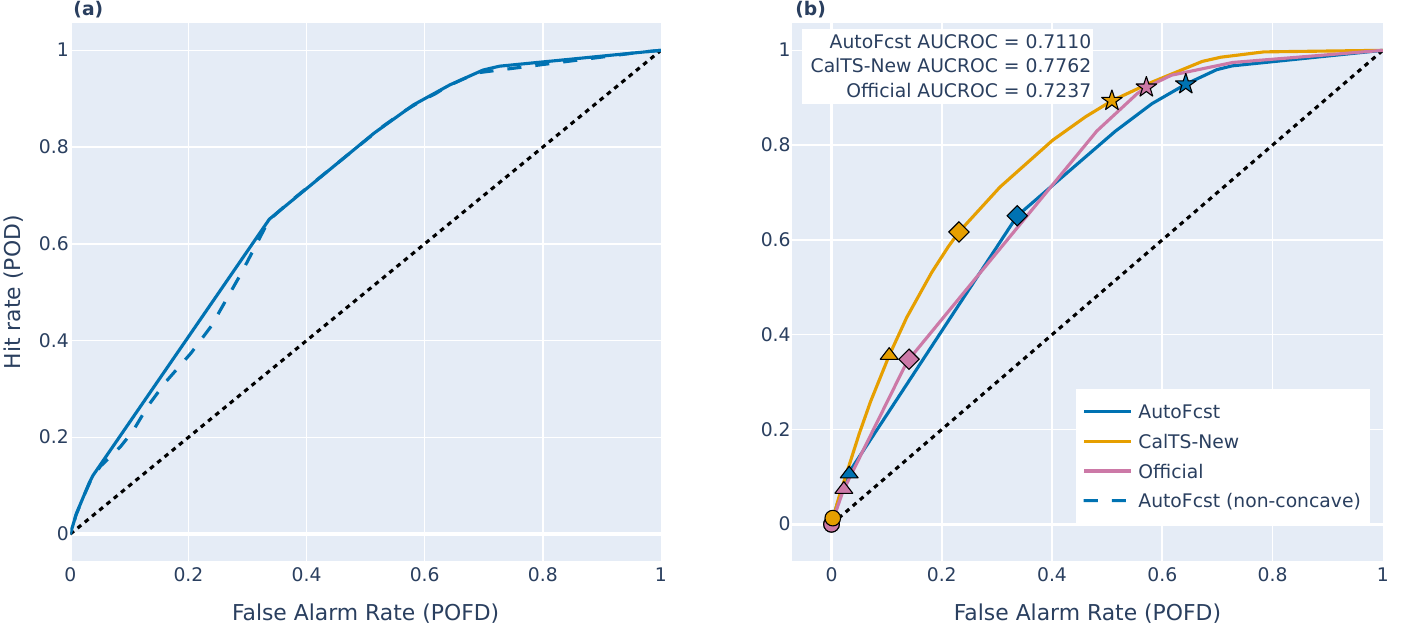}}
  \caption{(a) Curves for the standard ROC curve (dashed) and the concave ROC curve (solid) are displayed for the AutoFcst lead day 1 forecasts. The concave ROC curve is the smallest concave curve to the upper left of the ROC curve.(b) Concave ROC curves with the corresponding AUCROC for the lead day 1 AutoFcst, Official, and CalTS-New forecasts. Markers show the locations on the ROC curve of 10\% (star), 20\% (diamond), 30\% (triangle), and 40\% (circle) threshold probabilities.
}\label{fig:roc}
\end{figure}

To understand discrimination ability, it has been argued that ROC curves should ideally be concave\footnote{The literature varies in its usage of ``concave'' and ``convex'' when discussing ROC curves.  We use the term ``concave ROC curve'' to describe a ROC curve whose graph is concave, meaning that any line segment whose endpoints are on the curve lies on or below the curve. Notions of convexity and concavity regarding ROC curves are related through the following property: The area under a concave ROC curve, obtained by first recalibrating the forecast, is the area of the convex hull of the set of points from the original ROC curve with the addition of the point (1, 0).} but are only sometimes so with empirical data \citep{pesce2010convexity, gneiting2022receiver}. This is because assessing discrimination (or potential predictive performance) relies on the assumption that higher forecast probabilities imply higher event probabilities. Since potential predictive performance does not rely on calibration, one can produce concave ROC curves by applying a conditional bias correction via isotonic regression \citep{fawcett2007pav}. Visually this corresponds to taking the smallest concave curve to the upper left of the ROC curve. An example of the difference between concave and non-concave ROC curves is shown in Fig.~\ref{fig:roc}a.

Often the area under the ROC curve (AUCROC) is calculated \citep{Marzban_2004}. The AUCROC is a measure of potential predictive performance with a value of 1 indicating perfect discrimination ability and a value of 0.5 indicating no discrimination ability. Despite AUCROC sometimes being used in the literature as an overall predictive performance measure (see, e.g., the list in \citet{hand2013area}), such use can lead to misleading conclusions since it ignores forecast (mis)calibration \citep{hand2009measuring, hand2023notes}.  

Figure~\ref{fig:roc}b shows concave ROC curves for the three thunderstorm forecasts. CalTS-New clearly has the highest AUCROC with its concave ROC curve dominating the other two. Official has a slightly higher AUCROC than AutoFcst, with their ROC curves crossing several times. The ranking of discrimination ability based on AUCROC is consistent with the ranking of discrimination ability at lead day 1  based on the $\DSC$ component in Fig.~\ref{fig:brier_mcb_dsc}b and ~\ref{fig:brier_mcb_dsc}c. The Ideal, Over and Under synthetic forecast systems of Section~\ref{s:performance diagrams} have identical AUROC values, consistent with the fact they have identical discrimination ability, while the Jitter forecast system has lower AUCROC.

Consistent rankings of discrimination ability between AUCROC and DSC are not guaranteed. \citet{hand2009measuring} argues that AUCROC can give potentially misleading rankings if the ROC curves cross (e.g., the Official and AutoFcst ROC curves in Fig.~\ref{fig:roc}b) as it is possible for one forecast system to have a higher AUCROC than another, yet the latter may have better performance across the vast majority of threshold probabilities after both systems are calibrated. Furthermore, \citet{hand2009measuring} showed that when expressing AUCROC as a scoring rule in the form of Eq.~(\ref{eq:mixture representation}), the measure $H$ is dependent on the empirical data itself. This means that when comparing potential predictive performance of different forecast systems, AUCROC may apply a different weighting for each risk threshold for each forecast system. Nonetheless, when ROC curves for calibrated forecast systems don't cross, there exists a satisfying relationship between AUCROC, DSC, AUCPR and Murphy curves as detailed in Appendix~D.

We have discussed several measures of discrimination: AUCROC, DSC, AUCPR and maximum CSI\footnote{Generally speaking, POD and SR cannot be optimized simultaneously. One possible balancing act is to optimize the average of the two, and both being rates, a suitable average might be the harmonic mean of POD and SR (also known as the F1 score). One can show that the CSI is a function of the F1 score and that CSI and F1 are optimized simultaneously at the same threshold probability \citep{Hand_2021}. Maximum CSI is a measure of discrimination in this sense.}. All these measures may rank discrimination ability differently. AUCROC and $\DSC$ answer two different questions. For a sequence of forecast-observation pairs, the AUCROC is the probability that a random event taken from that sequence had a higher probability predicted by the forecast system than the predicted probability of a random non-event taken from the sequence. It is independent of the choice of proper scoring rule $S$. $\DSC$, on the other hand, measures the improvement of the recalibrated forecast improves over the best constant reference forecast as measured by $S$. If one wants to control the relative importance of different risk thresholds then DSC, calculated using a suitably chosen proper scoring rule $S$, is likely to be the more appealing discrimination measure.

Finally, precision--recall curves may provide a better visual display of discrimination compared to ROC curves when datasets are imbalanced (e.g., when events are climatologically rare) \citep{Saito_2015} and one could consider using them or performance diagrams alongside ROC curves in the Triptych approach. If precision--recall curves are used, then it is important to be aware of interpolation methods for generating the ``achievable precision--recall curve'' (which is the analogue to the concave ROC curve \citep{davis2006relationship}), the ``unachievable region'' \citep{boyd2012unachievable}, and appropriate methods for estimating AUCPR \citep{Boyd_2013}.

\section{Evaluation of categorical thunderstorm outlooks using the FIRM score}

In this section, we illustrate how categorical forecasts can be compared against probability forecasts using proper scoring rules. Recall that Bureau categorical thunderstorm outlooks have three categories (nil thunderstorm, thunderstorm possible, thunderstorm likely) defined using the probability bins $[0, 9.5\%)$, $[9.5\%, 29.5\%)$ and $[29.5\%, 100\%]$. In this case, there is a mapping from a probability forecast to a categorical forecast, and the important feature for measuring forecast accuracy is the predictive performance for the two categorical thresholds $\theta_1=0.095$ and $\theta_2=0.295$. In this context, a categorical threshold can also be interpreted as a decision threshold, since the threshold determines what action to take  (i.e. which forecast category to issue). By taking the mixing measure $H$ of Eq.~(\ref{eq:mixture representation}) to be positive point mass with weights $w_1$ and $w_2$ concentrated on decision thresholds $\theta_1$ and $\theta_2$, one obtains the proper scoring rule
\begin{equation}\label{eq:firm score}
S(x,y) = w_1 S_{\theta_1}(x,y) + w_2 S_{\theta_2}(x,y),
\end{equation}
where $x$ is the forecast probability, $y$ is the observation, and $S_{\theta_i}$ is defined by Eq.~\ref{eq:elementaryscore}. A lower score is better. The mean score $\bar S$ is the weighted sum of the Murphy curve values at the thresholds $\theta_1$ and $\theta_2$. We call the scoring rule $S$ of Eq.~(\ref{eq:firm score}) the FIRM score, since it is a special case of the FIRM scoring framework introduced by \citet{taggart2022scoring}. The scoring rule $S$ of Eq.~(\ref{eq:firm score}) can be re-expressed as a scoring matrix for categorical forecasts, as shown in Table~\ref{scoringmatrix}. This provides a fair way to compare predictive performance of probability forecasts with categorical forecasts using proper scoring rules. 

\begin{table}[h]
\caption{FIRM scoring matrix for the three category thunderstorm forecasts based on Eq.~(\ref{eq:firm score})}\label{scoringmatrix}.
\begin{center}
\begin{tabular}{c|cccrrcrc}
\hline
Forecast category & $\text{Observed non-event}$ & $\text{Observed event}$\\
\hline
 Nil thunderstorm & 0 & $2(w_1(1-\theta_1) + w_2(1-\theta_2))$ \\
 Thunderstorm possible & $2w_1\theta_1$ & $2w_2(1-\theta_2)$  \\
 Thunderstorm likely & $2(w_1\theta_1 + w_2 \theta_2)$ & 0 \\
\hline
\end{tabular}
\end{center}
\end{table}

For a graphical interpretation of the FIRM score, one can look at the points on the Murphy diagram that correspond to each threshold $\theta_i$. For example, looking at the lead day 1 performance in Figs.~\ref{fig:murphy} and~\ref{fig:murphy_ci} for the three category case where $\theta_1=0.095$ (9.5\%) and $\theta_2=0.295$ (29.5\%), the ranking of Official and AutoFcst against each other will depend on the choice of weights $w_i$. However, if instead $\theta_1=0.095$ and $\theta_2=0.35$, then Official will have performed better than AutoFcst based on the FIRM score irrespective of the choice of weights $w_i$.

\subsection*{TS outlook results}\label{ts_outlooks}

While the Bureau's lead day 1 categorical thunderstorm outlooks were produced manually by the TSHR team, they could in theory be automatically generated from the AutoFcst or CalTS-New probabilistic forecasts. To assess how each of these system's outlooks would compare with the TSHR team's outlooks, we calculate the difference between mean FIRM scores, using equal weights ($w_i=1$) for both categorical decision thresholds of 9.5\% and 29.5\%. We also generate 95\% confidence intervals using the Diebold--Mariano test statistic. Table~\ref{firmdiff} presents the difference $\overline{\FIRM}_{\mathrm{TSHR}} - \overline{\FIRM}_{\mathrm{prob}}$ in mean FIRM scores for the TSHR thunderstorm outlooks and one of the other probabilistic forecast systems. Consequently, negative values in this table indicate that TSHR team outlooks have higher performance compared to the other forecast systems, while the positive values indicate the opposite.

\begin{table}[h]
\caption{Difference ($\overline{\FIRM}_{\mathrm{TSHR}} - \overline{\FIRM}_{\mathrm{prob}}$) in mean FIRM scores for the thunderstorm outlooks at lead day 1, with 95\% confidence intervals estimated using Diebold--Mariano test statistic.}\label{firmdiff}.
\begin{center}
\begin{tabular}{cccccc}
\hline
Forecast $\mathrm{TSHR}$ & Forecast $\mathrm{prob}$ & $\overline{\FIRM}_{\mathrm{TSHR}} - \overline{\FIRM}_{\mathrm{prob}} (\times 10^3)$ &  95\% Confidence intervals ($\times 10^3$)\\
\hline
TSHR team & Official & -57.10 &  (-74.67, -39.53) \\
& AutoFcst & -53.79 & (-71.24, -36.34) \\
& CalTS-New & 1.25 & (-14.96, 17.47) \\
\hline
\end{tabular}
\end{center}
\end{table}

The difference in mean FIRM scores show that the performance of the TSHR team is higher than that of both AutoFcst and Official. However, the difference between FIRM scores between the TSHR team and CalTS-New is marginal. In fact, CalTS-New outperforms both Official and AutoFcst as measured by this FIRM score, as can be seen from the Murphy diagrams presented in Figs.~\ref{fig:murphy} and~\ref{fig:murphy_ci} when considering decision thresholds of 9.5\% and particularly 29.5\%.

\section{Miscalibration-Discrimination decomposition of the FIRM scores}\label{mcb-dsc_FIRM}
\begin{figure}[h]
 \centerline{\includegraphics[width=42pc]{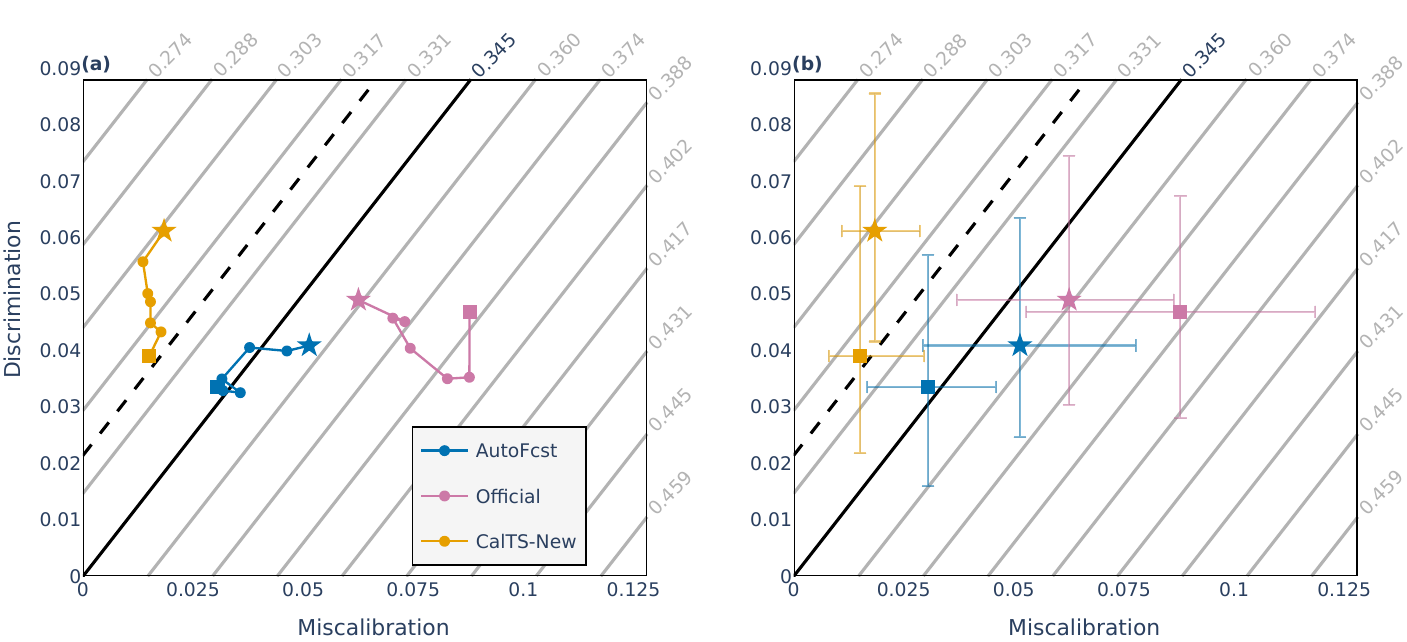}}
  \caption{As Fig.~\ref{fig:brier_mcb_dsc}, but for the FIRM score rather than the Brier score (diagonal gray lines correspond to the mean FIRM score). Confidence intervals were generated from 500 circular block bootstrap resamples.
}\label{fig:mcb_dsc_firm}
\end{figure}

Previously we constructed miscalibration--discrimination diagrams based on the Brier score, which weights all user decision thresholds equally (Fig.~\ref{fig:brier_mcb_dsc}). However, if we care primarily about the thresholds 9.5 and 29.5\%, then for our three forecast systems it is possible to perform the CORP decomposition (Eq.~(\ref{eq:mcbdsc_decomposition})) where $S$ is the FIRM score (Eq.~(\ref{eq:firm score})) with $w_1 = w_2 = 1$, $\theta_1 = 0.095$, and $\theta_2 = 0.295 $. To visualize these components, we plot them on a miscalibration--discrimination diagram in Fig.~\ref{fig:mcb_dsc_firm}.

CalTS-New has the lowest mean FIRM score, just as it also had the lowest mean Brier score. However, as noted previously, the ranking of forecast performance may depend on the choice of proper scoring rule, or equivalently on the choice of mixing measure $H$ in Eq.~(\ref{eq:mixture representation}). This can be observed when comparing Official with AutoFcst at lead day 1. When $S$ is the Brier score (where $H$ places equal weight on all decision thresholds), Official performed better than AutoFcst (Fig.~\ref{fig:brier_mcb_dsc}), while when $S$ is the FIRM score (where $H$ concentrates equal weight on two thresholds only), AutoFcst performs marginally better than Official (Fig.~\ref{fig:mcb_dsc_firm}).

CalTS-New has superior calibration based on the MCB component regardless of whether $S$ is the Brier or FIRM score. However, AutoFcst has a lower (better) MCB value than Official when $S$ is the FIRM score, while the opposite is true when $S$ is the Brier score. The differences in the MCB component between AutoFcst and Official are not statistically significant, but the point remains that the choice of scoring rule impacts the ranking of forecast reliability based on the MCB component. It is also possible for the ranking of discrimination ability to change based on the choice of FIRM score or Brier score for $S$ when calculating the DSC component. This was observed when the analysis was repeated for other parts of Australia (not shown). This implies that it is important to consider the choice of scoring rule $S$ through the construction of $H$, so that measures of performance, calibration, and discrimination align with how users utilize the forecasts.

\section{Conclusions}
Over the last decade there have been substantial theoretical and practical advances for diagnosing the quality of probabilistic forecasts for binary outcomes (probabilistic classifiers). Some of these have provided significant upgrades to tools such as reliability diagrams, while others have added new items (e.g. Murphy diagrams) to the toolkit. The application of isotonic regression for the recalibration of forecasts has given rise to stable reliability diagrams, the CORP decomposition of mean scores into miscalibration, discrimination and uncertainty components \citep{dimitriadis2021stable}, and the production of concave ROC curves and AUCROC statistics with better diagnostic properties than their traditional counterparts \citep{fawcett2007pav, pesce2010convexity, gneiting2022receiver}. The representation of proper scores (such as the Brier score) as an integral of elementary scores \citep{ehm2016quantiles} has given rise to diagnostics of predictive performance across the range of probabilistic decision thresholds, an interpretation of overall score as the weighted mean of economic loss over the range of decision thresholds, and a method for proper scoring rule selection given weights on decision thresholds suitable for the application at hand.

In this paper we have showcased these advances specifically for practitioners within the meteorological community by using forecasts from three probabilistic thunderstorm forecast systems at the Australian Bureau of Meteorology. Diagnostic tools illustrated include:
\begin{enumerate}
\item miscalibration--discrimination diagrams, which show overall predictive performance as well as overall reliability and discrimination information;
\item Murphy diagrams, which show predictive performance for each probabilistic decision threshold;
\item CORP reliability diagrams, which show calibration of a forecast system conditioned on the forecast, whilst avoiding problems arising from ad hoc binning choices; and
\item concave ROC curves, which provide one particular diagnostic of discrimination.
\end{enumerate}

A common thread in these advancements is that the methods have a sound theoretical basis for probabilistic forecasting, thereby avoiding misguided inferences and forecaster's dilemmas. We demonstrated that some interpretations of categorical performance diagrams that have gained currency within parts of the meteorology community can lead to the incorrect ranking of the overall predictive performance of forecast systems. Such inferential pitfalls should be avoided and instead if performance diagrams are used, they should be used as a diagnostic of discrimination ability.

While reliability diagrams, Murphy diagrams and concave ROC curves do not depend on the choice of proper scoring rule, overall ranking of forecasts as well as overall measures of miscalibration and discrimination from the CORP decomposition may depend on that choice. We have emphasized a paradigm where the proper scoring rule is selected by assigning appropriate weights to each probabilistic decision threshold. When equal weights are chosen, the Brier score emerges. For categorical products, like the Bureau's thunderstorm outlooks, positive weights are only assigned to the categorical thresholds, whereby the FIRM score emerges. We demonstrated that the FIRM score provides a fair way of comparing probabilistic and categorical forecasts for the generation of categorical products and illustrated that forecast rankings can indeed depend on score choice, again emphasizing the importance of appropriate score selection for the task at hand. This paradigm of score selection based on threshold weighting offers opportunities for social science and business research into methods to elicit from forecast users appropriate weights to put on decision thresholds.

Having applied these methods to Bureau thunderstorm forecasts, we have gained actionable insights that can be used to improve the thunderstorm forecast process and quality of its output. For example, we can advise Bureau meteorologists and decision makers which of the forecast systems provides the best starting point for the production of Official probabilistic gridded forecasts, and some simple manual edits that meteorologists can apply to conditionally bias correct that starting point. We can also advise which of the current manual edits appear to be adding value to the product, and which appears to be subtracting value. By identifying the highest performing recalibrated forecast system, as measured by the FIRM score, we can also make recommendations regarding the best starting point for the categorical thunderstorm outlook, and which (uncalibrated) thresholds to use to generate those categorical forecasts.

Apart from ROC curves, the methods and diagnostic tools recommended in this paper can also be applied to other types of single-valued forecasts, such as quantile forecasts (e.g. 90th percentile of precipitation in millimeters) \citep[e.g.,][]{gneiting2023model} and mean-value forecasts (e.g. expected temperature in degrees Celsius). We have made an Xarray-based, Python implementation of these methods publicly available at \url{https://github.com/nci/scores}. These methods can be used by the meteorological community for a user-centric approach to evaluate probabilistic and categorical forecasts.

\section*{Acknowledgments}
We thank Robert Warren, Callum Stuart, Elizabeth Ebert and Ryan Holmes from the Bureau of Meteorology, and three anonymous reviewers for providing constructive feedback on earlier drafts of this manuscript. We are also grateful to Robert Warren for providing us with climatological lightning data and forecast data from the new research version of Calibrated Thunder.
\section*{Appendix A}
\subsection*{Details of the construction of the climatological reference forecast}
As a benchmark reference forecast, we use an estimate of the daily climatological lightning probability at each grid point. The climatology was created by first counting the number of lightning strikes within a 10 km radius of each point for every day from 1 December 2014 to 30 November 2021. Similar to the observation grids, strike counts were then converted to a binary lightning flag and the resulting grids were averaged across the seven years to obtain the raw lightning probability for each day. Finally, spatial and temporal smoothing was applied to reduce fine-scale variability (caused by the short length of the climatology) using Gaussian filters with standard deviations of 40 km and 15 days.

\section*{Appendix B}
\subsection*{The CSI is not a proper scoring rule for probabilistic forecasts}
Consider a forecast system that issues probabilistic forecasts for binary outcomes. Suppose that $n$ forecast cases have already been made. Let $\mathrm{CSI}_{n,\theta}$ denote the CSI value for these $n$ forecast cases using the threshold probability $\theta$. Suppose that for the $(n+1)$th forecast case, a forecaster assesses that the probability of an event is $p$. With a little algebra, it can be shown that the expected value of $\mathrm{CSI}_{n+1,\theta}$ is maximized by forecasting an event if and only if
\[p \geq \mathrm{CSI}_{n,\theta} / (\mathrm{CSI}_{n,\theta} + 1)\]
\citep{mason1989dependence}. Now if $\theta > p \geq \mathrm{CSI}_{n,\theta} / (\mathrm{CSI}_{n,\theta} + 1)$, then the expected CSI can only be maximized by forecasting an event. However, an event can only be forecast by issuing a probability that is at least $\theta$. In other words, the forecaster cannot issue their true assessment $p$ whilst optimizing their expected CSI. Similarly if $\theta \leq p < \mathrm{CSI}_{n,\theta} / (\mathrm{CSI}_{n,\theta} + 1)$, then the expected CSI cannot be maximized by forecasting $p$. In these circumstances, the forecaster faces the dilemma of either issuing their best probabilistic assessment $p$ or a different probability that will maximize their expected CSI.

\section*{Appendix C}
\subsection*{Calculating proper scoring rules from the mixture representation}
Given a mixing measure $H$, we show how the mixture representation of Eq.~(\ref{eq:mixture representation}) can be used to calculate a formula for a proper scoring rule $S$ that is expressed in terms of the probabilistic forecast $x$ and binary outcome $y$ without reference to decision thresholds $\theta$.

Suppose that $H$ is the uniform measure, so that $\dd H(\theta) = \dd\theta$. In the case when $y=0$, Equations~(\ref{eq:mixture representation}) and (\ref{eq:elementaryscore}) give
\begin{align*}
S(x,0) 
&= \int_0^1 S_\theta(x,0)\,\dd\theta \\
&= \int_0^x 2\theta\,\dd\theta \\
&= x^2.
\end{align*}
On the other hand, when $y=1$, 
\begin{align*}
S(x,1) 
&= \int_0^1 S_\theta(x,1)\,\dd\theta \\
&= \int_x^1 2(1-\theta)\,\dd\theta \\
&= (x-1)^2.
\end{align*}
Putting both together yields $S(x,y)=(x-y)^2$, which is the Brier score.

A similar calculation using the measure $\dd H(\theta) = (2\theta(1-\theta))^{-1}\,\dd\theta$ yields the logarithmic score $S(x,y)=-y\log(x) - (1-y)\log(1-x)$. This is left as an exercise for the reader.

\section*{Appendix D}
\subsection*{The relationship between ROC, precision--recall diagrams,  performance diagrams, DSC, and Murphy diagrams}
Consider two calibrated (via isotonic regression if required) forecast systems A and B. Theorem 3 in \citet{dimitriadis2023evaluating} and Theorem 3.2 in \citep{davis2006relationship} imply the equivalence of the following statements:
\begin{enumerate}
\item The ROC curve for A lies on or above the ROC curve for B.
\item The precision--recall curve for A lies on or above the PR curve for B
\item The categorical performance diagram curve for A lies on or to the right of the categorical performance diagram curve for B.
\item The Murphy curve for A lies on or below the Murphy curve for B.
\item DSC for A is higher or equal to DSC for B, regardless of the choice of proper scoring rule $S$.
\end{enumerate}

The concave ROC curve for CalTS-New dominates the concave ROC curves of Official and AutoFcst in Figure~\ref{fig:roc}b. This means that the above statements are true with recalibrated CalTS-New as System A and recalibrated versions of either Official or AutoFcst as System B. Since the concave ROC curves of Official and AutoFcst cross, the above statements do not hold true when comparing calibrated versions of  those two forecast systems.
% Submissions are not required to reflect the precise reference formatting of the journal (use of italics, bold etc.), however it is important that all key elements of each reference are included.

\end{document}